\documentstyle [12pt]{article}
\begin{document}
lgonzalz@vxcern.cern.ch
~~~~~~~~~~~~~~~~~~~~~~~~~~~~~~~~~~~~~~~~~~~~~~~~~~~~~~~~~~~~~~~~~~~~~~~~March
1997
\vskip 2.5cm
\centerline {\bf LORENTZ INVARIANCE AND SUPERLUMINAL PARTICLES}
\vskip 2cm
\centerline {\bf L. GONZALEZ-MESTRES}
\vskip 5mm
\centerline {Laboratoire de Physique Corpusculaire, Coll\`ege de France}
\centerline {11 pl. Marcellin-Berthelot, 75231 Paris Cedex 05 , France}
\vskip 3mm
\centerline {and}
\vskip 3mm
\centerline {Laboratoire d'Annecy-le-Vieux de Physique des Particules}
\centerline {B.P. 110 , 74941 Annecy-le-Vieux Cedex, France}
\vskip 3cm
{\bf Abstract}
\vskip 4mm
If textbook Lorentz invariance is actually
a property of the equations describing a sector
of matter above some critical distance scale,
several sectors of matter with different
critical speeds in vacuum can coexist and an absolute rest frame (the vacuum
rest frame, possibly related to the local rest frame of the expanding Universe)
may exist without contradicting the apparent Lorentz invariance felt by
"ordinary" particles (particles with critical speed in vacuum equal to $c$ ,
the speed of light). Sectorial Lorentz invariance, reflected by the fact that
all particles of a given dynamical sector have the same critical speed in 
vacuum, will then be an expression of a fundamental sectorial symmetry
(e.g. preonic grand unification or extended supersymmetry) protecting a 
parameter of the equations of motion.
\vskip 3mm
We study the breaking of Lorentz invariance in such a scenario, with
emphasis on mixing between the "ordinary"
sector and a superluminal sector, and discuss with examples the consequences 
of existing data. The sectorial universality of the  
value of the high-energy speed in vacuum, even exact, does not necessarily 
imply that Lorentz invariance is not violated and does not by itself exclude 
the possibility to produce superluminal particles at accelerators or to find
them in experiments devoted to high-energy cosmic rays. Similarly, the
stringent experimental bounds on Lorentz symmetry violation at low energy
cannot be extrapolated to high-energy phenomena.
Several basic questions
related to possible effects of Lorentz symmetry violation are discussed,
and potential signatures are examined. 
\vskip 1cm
{\bf 1 . INTRODUCTION}
\vskip 5mm
In several previous papers [1 , 6] , we discussed possible properties of a new 
class of superluminal particles which, contrary to tachyons [7], have positive
masses and energies. Whereas the later respect Lorentz invariance and are
just space-like states of "ordinary" particles (particles with critical speed
in vacuum equal to $c$ , the speed of light), the new superluminal particles
would be expressions of new degrees of freedom of matter and have critical
speeds in vacuum substantially different from $c$ (in general, by several
orders of magnitude). They may obey sectorial Lorentz invariances, with a 
critical speed $c_i~\gg ~c$ playing the role of the space-time metric parameter
for the $i$-th superluminal sector. While absolute space and time can 
perhaps be defined in an absolute rest frame (the "vacuum rest frame"), the 
geometry of the space-time felt by matter depends on its dynamical properties.
Particles from different sectors, moving at a common velocity ${\vec {\bf v}}~
\neq ~0$ with respect to the vacuum rest frame (assuming such a frame exists),
will not feel the same space-time. Contrary to standard relativity,
Lorentz contraction, as seen in the vacuum rest frame, will be an absolute 
and physically meaningful phenomenon. It will depend on the critical speed of
the particle. At low energy, and as suggested by
experimental tests of Lorentz invariance, 
we expect superluminal particles to be very weakly coupled to
ordinary ones. In spite of this weak coupling, high-energy experiments 
(at accelerators or devoted to cosmic rays) can 
potentially discover the new particles (e.g. [5 , 6]) .
\vskip 4mm
If the critical speed in vacuum is not an absolute expression of any intrinsic
geometry of space-time, but depends (like the space-time geometry felt by the
particle) on the deep dynamical properties of the matter under consideration,
the fact that several particles have with great precision the same high-energy 
speed (i.e. the high $p$ limit of ${\mathbf \nabla }_{\vec {\mathbf p }}~E$ , 
where 
${\vec {\mathbf p }}$ stands for momentum, $E$ for energy and $p$ is the modulus
of ${\vec {\mathbf p }}$)  
in vacuum is by itself the signature of a fundamental symmetry of this 
sector of matter (like preonic grand unification or extended supersymmetry).
If, in the vacuum rest frame, free particles obey equations of the type:
\equation
-~A~\partial^2 \phi /\partial t^2~~+~~iB~\partial \phi /\partial t~~+~~\Sigma
_{j=1}^3~\partial ^2 \phi /\partial x_j^2~~-~~D~\phi ~~~=~~~0
\endequation
where $A$ , $B$ and $D$ are sectorial parameters and may be different 
for different particles,
the possible universality of the critical speed inside a dynamical sector
reflects an exact sectorial
symmetry of the parameter $A$ . The non-universality of mass inside 
the sector corresponds to a symmetry breaking effect acting on the parameter
$D$ . This would be perfectly compatible, for instance, with standard scenarios
of supersymmetry breaking where the spontaneous breaking is a low-energy 
effect and leaves physics unchanged in the high-energy region above the
symmetry-breaking scale. Thus, the apparent intrinsic properties of
space and time, as felt by ordinary matter
(experiment indicates that $B$ is small), would just reflect the fact that
the sectorial universality of $A$ (giving the high-energy speed of particles)
is preserved to a very good approximation
by symmetry-breaking phenomena. The exact 
universality of the high-energy speed would not by itself
imply exact sectorial Lorentz
invariance: $B~\neq ~0$ violates such an invariance, but as will be seen later
finite-energy effects or new Lorentz-violating terms can also arise. In what
follows, we assume for simplicity
that the parameter $B$ vanishes or can be neglected.
\vskip 6mm
{\bf 2 . LORENTZ SYMMETRY VIOLATION}
\vskip 5mm
Assume that: a) 
each sector has a conserved sectorial symmetry broken only spontaneoulsy
by a low-energy phenomenon (like standard breaking of GUTs or supersymmetry); 
b) all particles (or, at least, all basic constituents) belong to a single 
multiplet of the symmetry. 
If the interaction between two different sectors (the ordinary sector and
a superluminal one)
preserves both sectorial symmetries (but breaks the two Lorentz invariances),
several possible mechanisms can be considered  
which we describe in a simplified way:
\vskip 5mm
{\bf 2a. Intrinsic violations}
\vskip 5mm
A typical example is provided by terms due to a possible phase transition
at very high energy (see, e.g. [2 , 5]) above which Lorentz invariance, and
even possibly the particles under consideration, do no longer exist.
If $k_i$ is the critical wave vector scale of the 
$i$-th superluminal sector [2] , such terms can be represented by powers of
$k_i^{-2}~\Sigma _{j=1}^3~\partial^2 /\partial x_j^2$ times the usual laplacian.
For the ordinary particles, $k_i$ is to be replaced by  $k_0$ , where  $k_0$ is
the critical wave vector scale of the ordinary sector. To lowest order, in the 
$i$-th superluminal sector, we can add to (1) a term $\alpha
k_i^{-2}~(\Sigma _{j=1}^3~\partial ^2/\partial x_j^2)^2~ \phi $ , and similarly
for the ordinary sector with  $k_0$ instead of  $k_i$ . The sectorial parameter
$\alpha $ is expected to be of order 1 . In terms of wave function 
parameters, we can write in the vacuum rest frame from (1):
\equation
E~=~(2\pi)^{-1} h~A^{-1/2}~(k^2~+~D)^{1/2}
\endequation 
where $E$ is the energy, ${\vec {\mathbf k}}$ the wave vector
and we have taken $B~=~0$ in (1). The
inclusion of the new term would replace in (2) $k^2$ by 
$k^2~(1~+ \alpha k^2~k_i^{-2})$ if the sector is superluminal, and
by $k^2~(1~+ \alpha k^2~k_0^{-2})$ if we are considering
the ordinary sector. More generally, we can replace in (1) $k^2$ by
$k^2~f(\kappa )$ where $\kappa ~=~k^2~k_i^{-2}$ if we are dealing with a
superluminal sector and $\kappa ~=~k^2~k_0^{-2}$ in the ordinary sector.
We take $f(0)~=~1$ and expect $f(\kappa )$ to exhibit a singularity in the
region $\kappa \approx 1$ indicating the dissappearence of the Lorentz regime
and the transition to new expressions of fundamental vacuum dynamics.
If the sectorial symmetry is preserved at high energy, we expect the function
$f(\kappa )$ to be universal inside each sector and not to 
generate any difference
in critical speed between particles of the same sector. Terms involving higher
$t$ derivatives can be dealt with in a similar way.
\vskip 4mm
If the sectorial symmetry remains unbroken at high energy and
the value of
$A$ remains universal inside the sector under consideration, 
we do not 
expect obvious
measurable effects from conventional tests of Lorentz invariance.  
Contrary to the scenario recently considered by Coleman and Glashow [8] ,
the existence of very high-energy cosmic rays does not refute the dynamics we
propose. Photon decay, as well as Cherenkov radiation in vacuum by
ordinary charged particles, remain forbidden for very high-energy cosmic rays
as a consequence of the sectorial universality of the critical speed in vacuum.
At the highest energies observed ($E~\approx ~10^{20}~eV$ ,
$k~\approx ~10^{25}~cm^{-1}$), the ratio $k^2k_0^{-2}$ is of order $\approx
~10^{-16}$ (above the bound from [8])
for $k_0~\approx ~10^{-33}~cm$ . It becomes as small as $10^{-50}$
for $k~\approx ~10^8~cm^{-1}$ (atomic distance scale). 
If laboratory distance and 
time scales are fixed from phenomena occuring at atomic or larger
scales, the terms added to (1) do not seem to have a real influence on our 
natural definition
of space and time. If they ever become detectable, they will imply 
a definite failure of the relativity principle [9] , 
as it will become impossible to
write the same equations of motion in all inertial frames. A
more detailed discussion of 
high-energy phenomena in similar situations is given in Subsection 2b. 
\vskip 5mm
{\bf 2b. Mixing with superluminal sectors}
\vskip 5mm
In a previous paper [5] , we pointed out that the existence of superluminal
sectors of matter is not by itself in conflict with the basic principles
of standard cosmology [10] and quantum field theory (e.g. [11 - 13]) .
The new superluminal particles, which have positive masses and energies, are 
clearly different from the previously proposed tachyons 
[14 , 15] and have different
signatures. By definition, 
tachyons explicitly respect standard Lorentz invariance, whereas
our superluminal particles explicitly break it.
The superluminal sectors can couple to the ordinary one, but Lorentz symmetry
violations related to this mixing would be very difficult to observe, as can
be seen in two mixing schemes:
\vskip 5mm
a) {\bf Scheme 1 .} 
The two sectorial symmetries, as well as the two multiplets, are identical
(an isomorphism can be spontaneously generated). 
Then, a unique direct mixing scheme exists for particles of different sectors
and all ordinary particles undergo the same kind of mixing with the 
superluminal sector.
In the vacuum rest frame, an ordinary particle $\mid \phi _0> $
may mix with a superluminal one,  $\mid \phi _i> $ , through a 
hamiltonian $H$ such that:
\equation
<\phi _0\mid H \mid \phi _0>~~=~~c~(p^2~+~m_0^2~c^2)^{1/2}~~=~~E_0
\endequation
\equation
<\phi _i\mid H \mid \phi _i>~~=~~c_i~(p^2~+~m_i^2~c_i^2)^{1/2}~~=~~E_i
\endequation
\equation
<\phi _0\mid H \mid \phi _i>~~=~~<\phi _i\mid H \mid \phi _0>~~=~~
\epsilon~(p^2)
\endequation
where ${\vec {\mathbf p}}$ is the momentum, $p$ the modulus 
of ${\vec {\mathbf p}}$ , $m_0$ the mass of the ordinary particle, $m_i$
the mass of the superluminal particle and $\epsilon $ the momentum-dependent
energy mixing parameter. Zero masses are preserved if $\epsilon~(0)~=~0$ .
The eigenvectors of $H$ are:
\equation
\mid \phi >~~=~~(1~-~\alpha ^2)^{1/2}~\mid \phi _0>~+~ \alpha ~\mid \phi _i>
\endequation
where:
\equation
2~\alpha ~\epsilon~~=~~E_i~-~E_0~-~[(E_i~-~E_0)^2~+~4~\epsilon ^2]^{1/2}
\endequation
corresponding to the mixed ordinary particle,
with eigenvalue $E$ given by:
\equation
2E~~=~~E_0~+~E_i~-~[(E_i~-~E_0)^2~+~4\epsilon ^2]^{1/2}
\endequation
and
\equation
\mid \phi '>~~=~~(1~-~\alpha ^2)^{1/2}~\mid \phi _i>~-~  \alpha ~\mid \phi _0>
\endequation
which corresponds to the mixed superluminal particle,
with eigenvalue $E'$ satisfying:
\equation
2E'~~=~~E_0~+~E_i~+~[(E_i~-~E_0)^2~+~4\epsilon ^2]^{1/2}
\endequation
\par
In the above expressions for $E$ and $E'$ , the value of the square root is
to be taken with the same sign as $E_i~-~E_0$ . 
The speed of particle $\mid \phi >$ is given by:
\equation
{\vec {\mathbf v}}~({\vec {\mathbf p}})~=~
{\mathbf \nabla }_{\vec {\mathbf p }}~E~=~{\vec {\mathbf p}}~[\mu (p^2)]^{-1}
\endequation
where $\mu (p^2)$ is an effective mass given by:
\equation
[\mu (p^2)]^{-1}~~=~~2~dE/dp^2
\endequation
\par
$\epsilon (p^2)$ breaks Lorentz symmetry and reflects unknown dynamics.
If $\epsilon (p^2)~=~a~p~+~g~(p^2)$ where $p^{-1}~g~(p^2)~\rightarrow ~0$ as
$p^2~\rightarrow ~\infty $ , the high-energy symmetry would require the value
of $a$ to be universal inside the sector, whereas $g~(p^2)$ can undergo the
effect of low-energy symmetry breaking. Again, very high-energy cosmic rays
will not exhibit phenomena like photon decay or Cherenkov emission in vacuum 
by charged particles. The high-energy speed of all particles in the mixed
"ordinary" sector, as measured in the vacuum rest frame, will tend to 
$v_{he}$ , where $2~v_{he}~=~c~+~c_i~+~[(c~-~c_i)^2~+~4a^2]^{1/2}$ . 
If $c_i~\gg
~c$ and $a~\ll ~c_i^2$ , we get $v_{he}~\simeq ~c~-~a^2c_i^{-1}$ 
where $c^{-1}~(v_{he}~-~c)$ is of order 
$\approx 10^{-2}$ for $a~\approx ~10^{2}c$ and
$c_i~\approx 10^{-6}~c$ .
Scenarios involving such figures would be
excluded according to recent work by Coleman and Glashow [8] if,
contrary to the case we are considering, the shift in critical 
speed were not the same for all particles of the ordinary sector. 
Thus, the high-energy symmetry can play a crucial physical role
which deserves further study.
\vskip 4mm  
For our laboratory definition of space and time, we should
use the low-energy expansion of $\epsilon ~(p^2)$ . Low-energy effects can be
much smaller than high-energy effects if there is some high-momentum
threshold contained in the $p$-dependence of $\epsilon (p^2)$ . 
Assume that: a) $a~\approx ~10^{2}~c$ and
$m_i~c_i^2~\approx 1~TeV$ ;
b) between atomic scales and high energies,
$\epsilon (p^2)$ can be approximated by the expression:
$\epsilon ~(p^2)~\approx ~ a~p^3~(M^2c_i^{2}~+~p^2)^{-1}$ 
where $M_ic_i~\approx ~1~TeV/c$ is the low-momentum cut-off. At atomic scales, 
the energy mixing parameter $\epsilon$ 
will be of order $\approx 10^{-13}~eV$
and we get
a relative effect of order $\epsilon ^2~(m_0~c_0^2~m_i~c_i^2)^{-1}~ 
\approx 10^{-44}$ on the electron wave equation. Similarly, a free photon with
the same values of $\epsilon ~(p^2)$ and  $m_i~c_i^2 $ would undergo an
energy shift of $\approx 10^{-38}~eV$ , which amounts at atomic scales to
$\approx 10^{-41}$ times its energy. This does not in principle correspond
to a mass, which may be even smaller, but is already
below the most stringent experimental data: the $p$-dependence of $\epsilon $
can be made much weaker between the low-energy region and the cut-off.
With the previous numbers, the photon 
velocity shift would be $\approx 10^{-41}~c$ 
and consequently, according to
formulae from [6] , we expect anisotropies in the value of the
speed of light measured on earth to be 
below $\approx 10^{-44}~c$ if the vacuum rest frame is close to that suggested 
by cosmic microwave background radiation. Part of the effect will be washed 
out by the redefinition of space-time automatically performed by apparatus 
made of mixed ordinary
matter; this redefinition will in general depend on the material the apparatus
is made of. The laws of physics for the mixed ordinary sector will, at such 
precision levels, depend on the laboratory velocity with respect to the 
vacuum rest frame.
The mixing
with the superluminal particle at atomic scales would be $\approx 10^{-25}$ 
and, if the interaction with the superluminal sector occurs only through 
direct mixing of single particles, the superluminal contribution to an
electromagnetic vertex of mixed "ordinary" particles can be suppressed
by an extremely small factor. For all the discussed low-energy parameters, 
experiment permits much higher values than those obtained with 
our parametrization.
\vskip 4mm
Thus, the space-time geometry felt by low-energy systems made of 
(mixed) ordinary matter can show extremely small deviations from standard
Lorentz invariance, even if a more important effect arises in the kinematics
of very high-energy cosmic rays. It must also be realized that, 
even in the case 
of a small
breaking of the high-energy symmetry, finite-energy effects can partially 
protect high-energy cosmic rays. For instance, with
the above parametrization, a 10$^{13}$ eV photon would have an anomaly in the
vacuum-rest-frame $E/p$ ratio:
\equation
E/p~-~c~~\simeq ~~-~a^2~c_i^{-1}~(1~-~2p^{-2}~c_i^2~M^2)
\endequation
and, with the above parameters, $p^{-2}~c_i^2~M^2~
\approx ~10^{-2}$ at $E~\approx 10^{13}~eV$ . This term increases the
energy/momentum ratio as the momentum is lowered, and ill severely limit the 
phase space for a photon to decay into an electron-positron pair 
in the presence of a very small 
breaking of the high-energy symmetry. The effect will be even stronger if 
the momentum cutoff is higher than considered in our example, and may  
completely protect high-energy cosmic rays up to energies much higher than
those observed. 
Differences in speed will appear below the 
high-energy limit due to the finite-energy 
breaking of the high-energy symmetry, but will not necessarily prevent
the existence of high-energy cosmic rays.
The argument by Coleman and Glashow 
must be reconsidered in our scheme and, even in the presence of an 
asymptotic or finite-energy breaking of
the high-energy symmetry, it leads to less stringent bounds, 
not better than $\delta (E/p)$ (difference in $E/p$ ratio) $<~10^{-11}$ 
obtained assuming that a $10^{20}~eV$
proton exchanges all its momentum with another
particle. Bounds on $\delta (dE/dp)$ will be much weaker.
Breaking of the speed symmetry at high energy,
even by a finite energy effect, may have another
important consequence: not only 
particles which are stable at low momentum (in 
the vacuum rest frame) may be unstable at high momentum (as
pointed out in [8] for the photon), but conversely unstable particles may
become stable when accelerated or produced at high momentum. This may have 
cosmological implications and eventually be tested by high-energy experiments.
\vskip 4mm
At $p~\approx ~1~TeV/c$ and with the above figures, we have 
$\epsilon ~(p^2)~\approx ~10^{-2}~TeV$ and a mixing 
parameter $\approx ~10^{-4}$
between ordinary and superluminal particles, which means that the probability
for two mixed quarks with $TeV/c$ momenta
to simultaneoulsy turn into superluminal particles
is $\approx ~10^{-16}$ (not necessarily negligible at LHC luminosities).
Similarly, the probability for a hard virtual particle produced at LHC to
become superluminal would be  $\approx ~10^{-8}$ , which again cannot be
entirely disregarded. Apart from direct production of superluminal particles,
precision speed measurements for muons and protons at LHC energies
deserve consideration. Finally, it may be worth noticing that a mixing 
$\alpha ~ \approx ~10^{-4}$ 
in the ordinary sector would lead to interaction probabilites slightly lower
than expected, typically by a fraction $\approx ~10^{-3}$ , and similarly
for energy loses. In the presence of interference phenomena, the effect may
become larger.
\vskip 4mm
b) {\bf Scheme 2 .}
If symmetries or multiplets are different and 
direct mixing between single particles from different 
sectors is forbidden, mixing between the two sectors can still arise, for
instance, through insertions of pairs of superluminal lines inside 
self-energy graphs. In this case, the energy of an ordinary particle
can be written in terms of its momentum as:
\equation
E~(p)~~=~~(p^2~+~m^2)^{1/2}~+~\Sigma ~(p^2)
\endequation
where the first term ignores the mixing with the superluminal sector and
$\Sigma ~(p^2)$ contains the superluminal insertions. As before, we can write
$\Sigma ~(p^2)~=~s~p~+~t~(p^2)$ , where $p^{-1}~t~(p^2)~\rightarrow ~0$ as
$p~\rightarrow ~\infty $ , and (using the high-energy symmetry) require the
value of $s$ to be universal inside the sector while $t~(p^2)$ can reflect the
low-energy symmetry breaking. 
\vskip 4mm
Arguments similar to scheme 1 can then be developed, and again 
with the same philosophy we may expect very
small numbers in direct
searches for Lorentz invariance violations at low
energy even if high-energy effects turn out to be detectable. A particularly 
interesting possibility would be that the Higgs boson couples to a superluminal
scalar through a $\phi ^4$ term in the lagrangian (the squared modulus of
the Higgs field times the squared modulus of the superluminal scalar field
like in [1]),
and that the superluminal boson has a rest energy in the $TeV$ range.
Then, the new physics expected at $TeV$ scales on the grounds of Higgs theory
considerations could directly involve the production of pairs of superluminal
particles, together with other departures from the standard model. 
\vskip 6mm
{\bf 3. DIRECT SEARCH FOR SUPERLUMINAL PARTICLES}
\vskip 5mm
With the same figures as before, LHC collisions can potentially
involve interactions between superluminal components of the high-energy
particles. To the already specified suppression factors, 
we may be led to add phase space limitations.
If, for instance, $c_i~\approx ~10^6~c$ as before, naive on-shell
kinematics 
would require the momenta of the incoming quarks to cancel with $10^{-6}$
precision if superluminal particles must emerge from the collision
(otherwise, there would not be enough energy left to produce the free
superluminal particles). We would then expect
a strong extra supression factor from these considerations, 
although our lack of knowledge of the dynamics of superluminal particles
prevents us from formulating definite conclusions (for instance,
some dramatic effect might
be generated by a pair of strongly off-shell superluminal particles).
On the other hand, 
going beyond point-like field theory, it should be noticed that 
Lorentz contraction of ordinary particles may modify their direct mixing with
the superluminal sector (see [6] on the absolute physical meaning of Lorentz
contraction and its matter-dependence), as it modifies the relative
longitudinal
size between ordinary and superluminal particles. If superluminal particles
are smaller by orders of magnitude than ordinary ones (which would most likely
occur if $k_i~\gg ~k_0$), Lorentz contraction of ordinary particles
may lead to further enhancement 
of direct coupling between particles from the two sectors. Such effects are
new with respect to coventional particle physics, where Lorentz contraction
is a relative phenomenon.
The situation would even become
more favourable in future generations of higher-energy accelerators.
In the case of direct mixing,
the mixing parameter $\alpha $ will reach its asymptotic value only
well above the rest energy $m_ic_i^2$ of the superluminal particle
under consideration. In any case, it seems that some scenarios with
superluminal particles can be explored by LHC experiments, mainly
for values of $c_i$ between $c$ and $10^6~c$ . 
We shall not attempt to give here
a general parametrization, but obvious examples can be built, especially
for critical speeds in vacuum one or two orders of magnitude above $c$ .
Beam quality will play a crucial role to precisely
define these potentialities.
\vskip 4mm
High-energy cosmic rays of the superluminal sector under consideration
would satisfy 
modified versions of the GZK cutoff [16] allowing for higher energies and
larger source distances, and possibly reach cosmic ray detectors on earth,
like AMANDA [17]. Particles from other superluminal sectors, more weakly
coupled to ordinary matter, would be much less contrained by the GZK cutoff.
The main limitation for a superluminal cosmic ray to be able to
reach earth is likely to come from "Cherenkov" radiation 
(i.e. the emission of particles belonging to a sector with lower critical
speed) in vacuum.
However, if superluminal matter is very abundant in the Universe, 
this abundance
may generate other cutoffs of the GZK type for superluminal particles.
The superluminal particles possibly produced at accelerators will in general
not be the same that we can observe in cosmic ray detectors, especially
if several superluminal sectors of matter exist. The fact that very 
high-energy ordinary particles may undergo mixing with superluminal sectors is 
potentially a motivation for detailed measurements (e.g. speed, "Cherenkov"
emission in vacuum even at small rates...) in high-energy cosmic-ray
detectors.
\vskip 4mm
We therefore claim that the impressive results of low-energy searches for
Lorentz symmetry violation at low energy, excluding such phenomena down
to very small numbers (e.g. [18 , 19]), do 
really not allow to conclude against the
potentialities of the search for superluminal particles in high-energy
experiments. On the contrary, high-energy experiments may indeed be the only
way to find evidence for Lorentz symmetry violation, as well as for the
existence of superluminal sectors. As stressed in our previous papers
(e.g. [1, 5]), the observed stability of Lorentz invariance in
the low-energy region and the possible discovery of superluminal particles
in high-energy experiments are not really incompatible, since Lorentz 
symmetry violation would be essentially related to phenomena occuring at 
very high energy and very short distance scales. Massless 
ordinary particles may 
even have different low-energy and high-energy speeds
in vacuum, as discussed above.
\vskip 4mm
Signatures in high-energy experiments are a crucial point: can we identify
superluminal particles in such experiments? Cherenkov radiation in vacuum
(e.g. the emission of ordinary particles by the produced superluminal ones)
may be a crucial signature in many cases. At accelerators, its 
structure and energy
dependence would be easy to identify. In cosmic ray experiments, it may 
be detectable even if the particle couples very weakly to ordinary matter.
Ordinary cosmic rays resulting from this emission can reach the detector, even
if the superluminal particle does not. Then, the event could most likely not
by explained in terms of known astrophysical sources. 
In all cases, because of the very high energy/momentum ratio 
in the new dynamical sector, "Cherenkov"
radiation in vacuum
by superluminal particles will present special features. At an angle
of $\pi /2$ with respect to its momentum, a relativistic
superluminal particle (i.e. at speed $v_i~\sim ~c_i$) with momentum $p_i$
and energy $E_i$ in the vacuum rest frame can emit
an ordinary particle with energy $E~\approx E_i~c^2~c_i^{-2}$ and momentum 
$p~\approx E_i~c~c_i^{-2}~\simeq p_i~c~c_i^{-1}$ . This is the highest
possible momentum of an emitted single particle. The longitudinal momentum
of emitted single particles, with respect to the direction of motion of the
superluminal particle, is strongly limited. Cherenkov radiation in vacuum 
will most likely
be mainly made of very heavy virtual particles quickly decaying into
pairs of jets (hadronic or leptonic) back-to-back in order to account for the
small momentum left. Conversion of these vacuum-rest-frame figures to the
laboratory rest frame can be made according to formulae developed in [6] .
\vskip 4mm
Other signatures exist
as well
and can be explored. As stressed in previous papers [5 , 6] , if a high-energy
superluminal cosmic ray undergoes
a hard interaction inside the detector releasing a sizeable part of its energy,
we expect in most cases the production of two jets "back-to-back" due to
energy and momentum conservation. Furthermore, the event distribution would be 
nearly isotropic as a result of the relation between the energy and momentum
of the superluminal particle. Such a signature would be unambiguous, even at
very small event rate. Similarly, precise timing in accelerator experiments
may provide dramatic signatures [6]. Absolute
time resolution below $10^{-8}~s$ would in principle allow to directly notice
a speed well above $c$ in a $10~m$ detector.
\vskip 6mm
{\bf 4. NEUTRINO AND OSCILLATION SIGNATURES}
\vskip 5mm
As previoulsy suggested in our papers [1 - 6] , and discussed by Coleman and 
Glashow in their recent study [8] , a small violation of Lorentz invariance
may lead to observable effects in neutrino physics. The numbers obtained 
above do not allow,
in our scenario, to exclude Lorentz symmetry violation from low-energy 
neutrino data, but make further 
experiments particularly relevant and necessary. Furthermore, other particles
than neutrinos
can undergo velocity oscillations, especially at 
high energy, if such oscillations are allowed. 
\vskip 4mm
Due to the
existence of a preferred frame and to the intrinsic meaning of velocity with
respect to this frame, there is a real physical difference between oscillations
and high and low momenta. A high-energy particle is 
intrinsically different from a low-energy particle: no Lorentz transformation
can make them completely equivalent.
Exclusion plots for oscillations should include momentum-dependence
if Lorentz invariance is broken,
as they test different physical objects when the 
speed is modified. While low-energy and nuclear physics 
experiments can test important aspects of neutrino oscillations, 
including Lorentz invariance, these phenomena are not necessarily the same
that can be potentially uncovered by high-energy experiments.
Also, in scenarios like that of 
Subsection 2b , Scheme 1 , 
the global high-energy symmetry is preserved but gauge
symmetries become sectorial and approximate when mixing between the two
sectors is introduced. Therefore, high-energy experiments may indeed be able
to test new physics related to this mixing.
\vskip 4mm
If the above high-energy symmetry undergoes a small breaking leading to 
differences in high-energy speeds escaping the Coleman-Glashow constraints
(e.g. Subsection 2b , Scheme 1), the experimental
consequences can be important even if very high-energy cosmic rays 
(e.g. protons and photons) remain
allowed. At TeV energies, and assuming masses to be small enough (typically,
$\delta m^2~c^5~sin2\theta _m \ll E^2~\delta v~sin2\theta _v$ , where 
$E$ is the energy , $m^2$ the difference in squared mass, $\theta _m$
the mass mixing angle and  $\theta _v$ the velocity mixing angle [8]),  
with a velocity symmetry breaking $\delta v~v^{-1}~\approx ~10^{-19}$ ,
oscillation would
occur at the scale of the detector size for all produced ordinary particles
able to oscillate. Well above this value of $\delta v~v^{-1}$ , the mixing
would be full.
Such considerations are not only relevant for neutrinos: for 
instance, mixing with the 
superluminal sector at high momentum may result in high-energy mixing
between charged leptons. In this case,
the oscillation will depend on the details of
the $p$-dependence of the mixing (which reflects in turn unknown details of
dynamics inside the superluminal sector). 
Again, the mixing can be very small at low energy but significant 
at high energy, similar to Subsection 2b . LHC and cosmic ray
experiments may be able to put relevant bounds
on such phenomena. 
\vskip 1cm 
{\bf References}
\vskip 5mm
\par
\noindent
[1] L. Gonzalez-Mestres, "Properties of a possible class
of particles able to travel faster
than light", Proceedings of the Moriond Workshop on "Dark Matter in Cosmology,
Clocks and Tests of Fundamental Laws", Villars (Switzerland), January 21-28
1995 , Ed. Fronti\`eres, France. Paper astro-ph/9505117 of electronic library.
\par
\noindent
[2] L. Gonzalez-Mestres, "Cosmological implications of a possible class of
particles able to travel faster than light", Proceedings of the Fourth
International Workshop on Theoretical and Experimental Aspects of
Underground Physics, Toledo (Spain) 17-21 September 1995~, Nuclear Physics B
(Proc. Suppl.) 48 (1996). Paper astro-ph/9601090 .
\par
\noindent
[3] L. Gonzalez-Mestres, "Superluminal matter and high-energy cosmic rays",
May 1996 . Paper astro-ph/9606054 .
\par
\noindent
[4] L. Gonzalez-Mestres, "Physics, cosmology and experimental signatures of
a possible new class of superluminal particles", to be published in the
Proceedings of the International Workshop on the Identification of Dark
Matter, Sheffield (England, United Kingdom), September 1996 . Paper
astro-ph/9610089 .
\par
\noindent
[5] L. Gonzalez-Mestres, "Physical and cosmological implications of a possible
class of particles able to travel faster than light", contribution to the
28$^{th}$ International Conference on High-Energy Physics, Warsaw July 1996 .
Paper hep-ph/9610474 .
\par
\noindent
[6] L. Gonzalez-Mestres, "Space, time and superluminal particles", February
1997 . Paper mp$_-$arc 97-92 and physics/9702026 .
\par
\noindent
[7] See, for instance,
"Tachyons, Monopoles and Related Topics", Ed. by E. Recami,
North-Holland 1978 , and references therein.
\par
\noindent
[8] S. Coleman and S.L. Glashow, "Cosmic ray and neutrino tests of
special relativity", Harvard preprint HUTP-97/A008 , paper 
hep-ph/9703240 . 
\par
\noindent
[9] 
The relativity principle was formulated by
H. Poincar\'e, Speech at the St. Louis International Exposition of 1904 ,
The Monist 15 , 1 (1905).
\par
\noindent
[10] See, for instance, P.J.E. Peebles, "Principles of Physical Cosmology",
Princeton Series in Physics, Princeton University Press 1993 .
\par
\noindent
[11] S.S. Schweber, "An Introduction to Relativistic Quantum Field Theory",
Row, Peterson and Company 1961 .
\par
\noindent
[12] See, for instance,
R.F. Streater and A.S. Wightman, "$PCT$, Spin and Statistics, and All
That", Benjamin, New York 1964 ; R. Jost, "The General Theory of
Quantized Fields", AMS, Providence 1965 .
\par
\noindent
[13]  C. Itzykson and J.B. Zuber, "Quantum Field Theory", McGraw-Hill 1985 .
\par
\noindent
[14] See, for instance, E. Recami in [7] .
\par
\noindent
[15] See, for instance, E.C.G. Sudarshan in [7] .
\par 
\noindent
[16] K. Greisen, Phys. Rev. Lett. 16 , 748 (1966); G.T. Zatsepin and V.A. 
Kuzmin, Pisma Zh. Eksp. Teor. Fiz. 4 , 114 (1966).
\par
\noindent
[17]  
See, for instance, L. Bergstrom et al., "THE AMANDA EXPERIMENT: status and
prospects for indirect Dark Matter detection", same Proceedings as for ref.
[4]. Paper astro-ph/9612122 .
\par
\noindent
[18] S.K. Lamoreaux, J.P. Jacobs, B.R. Heckel, F.J. Raab and E.N. Forston,
Phys. Rev. Lett. 57 , 3125 (1986); D. Hils and J.L. Hall, Phys. Rev. Lett.
64 , 1697 (1990).
\par
\noindent
[19] M. Goldhaber and V. Trimble, J. Astrophys. Astr. 17 , 17 (1996).
\end{document}